\documentclass[fleqn,twoside]{article}
\usepackage{espcrc2}
\usepackage[dvips]{graphicx}
\usepackage[figuresright]{rotating}
\usepackage{amssymb}
\usepackage{textcomp}

\newcommand{\E}{E_{0}}
\newcommand{\R}{R_{0}}
\newcommand{\X}{X_{0}}
\newcommand{\LambE}{\lambda_{\mathrm{e}}}
\newcommand{\LambM}{\lambda_{\mathrm{m}}}
\newcommand{\AttenUnit}{\mathrm{g} / \mathrm{cm}^{2}}
\newcommand{\DepthUnit}{\mathrm{g} \cdot \mathrm{cm}^{-2}}

\newcommand{\AmS}{{\protect\the\textfont2
  A\kern-.1667em\lower.5ex\hbox{M}\kern-.125emS}}

\title{Spectrum of cosmic rays with energy above $10^{17}$~eV}

\author{S.P.~Knurenko
        \address[YKT]{Yu.G.Shafer Institute of Cosmophysical Research and Aeronomy,
		\\31 Lenin Ave., 677980 Yakutsk, Russia},
        V.P.~Egorova\addressmark[YKT]{},
        A.V.~Glushkov\addressmark[YKT]{}, 
        A.A.~Ivanov\addressmark[YKT]{},
		A.D.~Krasilnikov\addressmark[YKT]{},
		I.T.~Makarov\addressmark[YKT]{},
		A.A.~Mikhailov\addressmark[YKT]{},
		V.V.~Olzoyev\addressmark[YKT]{},
		M.I.~Pravdin\addressmark[YKT]{},
		A.V.~Sabourov\addressmark[YKT]{},
		I.Ye.~Sleptsov\addressmark[YKT]{},
        G.G.~Struchkov\addressmark[YKT]{}.
		}

\begin{document}

\begin{abstract}
There are some discrepancies in the results on energy spectrum from Yakutsk, AGASA,
and HiRes experiments. In this work differential energy spectrum of primary cosmic
rays based on the Yakutsk EAS Array data is presented. For the largest events values
of $S_{600}$ and axes coordinates have been obtained using revised lateral
distribution function. Simulation of converter's response at large distances showed
no considerable underestimation of particle density. Complex shape of spectrum in
region of $\E > 10^{17}$~eV is confirmed. After adjustment of parameters and
additional exposition at the Yakutsk array there are three events with energy $\E >
10^{20}$~eV.
\vspace{1pc}
\end{abstract}

\maketitle

\section{INTRODUCTION}

Research into the cosmic rays above $10^{17}$~eV spectrum shape and into intensity in
steep region near $10^{20}$~eV predicted by Greisen~\cite{bib1}, Zatsepin and
Kuzmin~\cite{bib2} are of great importance for their sources detection. Results obtained
in various experiments~\cite{bib3,bib4,bib5} differ from each other by factor 2 and more
in absolute intensity, but their shapes are similar. Variations of intensity cannot be
described with simple power law. At energies greater than the GZK--cutoff results are
inconsistent. At HiRes there is only one event with $\E > 10^{20}$~eV and the spectrum is
cut--off. AGASA has registered 11 such events ($\theta < 45^{\circ}$), this could be an
evidence for absence of a cutoff.

At the Yakutsk array after recent analysis have been carried out~\cite{bib3} there is only
one event with energy estimated to be greater than $10^{20}$~eV. To explain this
contradiction with AGASA A.~Watson assumed that at the Yakutsk array such showers are
skipped due to inadequate short integration time for large distances from the axis. In
this work we have studied affect of particle arrival time distribution at different
distances on estimated density for Yakutsk and AGASA. Simulation showed that distortion of
estimated density at the Yakutsk array cannot result in such miscalculation. More
substantively is the lateral distribution function (LDF) used for axis determination. We
provided axes coordinates determination with adjusted LDF for the largest events, which in
average led to increase of $S_{600}$. Besides, February 18 2004 a new event with energy
$10^{20}$~eV was detected at the Yakutsk array. As a result of these factors there are
three events of $\E > 10^{20}$~eV registered at the Yakutsk array.

\section{DENSITY MEASUREMENT AT LARGE DISTANCES FROM THE AXIS}

A shower disk at lagre distance from the axis ($R > 1500$~m) is quite thick. At Yakutsk
array and at AGASA a nearly similar logarithmic RC--converters of the signal from
photomultiplier tube (PMT) to digital code with $r \sim 10$~mcsec are used. At the Yakutsk
array for an event to be treated, a coincidence of signals from both detectors within
2~mcsec is required. Herewith input of converters is closed in 2~mcsec after coincidence.
In the case when shower front is wide, this may result in underestimation of the density.
Besides, there is a possibility, that station doesn't operate due to large difference in
particles arrival times on different detectors and so some showers might be skipped. These
circumstances have been pointed out by Watson in his report~\cite{bib6}. At AGASA input is
permanently open and in the case of wide signal this may lead to density overestimation
due to features of converter.

To examine the influence of the effects mentioned above, we have provided simulation of
detector's response for distances $R = 1050, 1500$ and 2000~m, based on the particle
distribution approximation obtained at AGASA~\cite{bib7}. A coefficient $K_{R}$ was
considered --- a ratio between density estimated with RC--convector and the one set with
program. For the Yakutsk array, in the case of detectors with large area, registering
large particle densities we got following values: $K_{1050} = 1.05$, $K_{1500} = 0.994$,
$K_{2000} = 0.76$. Same points for AGASA: $K_{1050} = 1.065$, $K_{1500} = 1.11$, $K_{2000}
= 1.2$. At 2000~m distance for Yakutsk there is 25\% underestimation, for AGASA --- 20\%
overestimation.

For the real experiment for the shower with $\E = 10^{20}$~eV at $R = 2000$~m about 2
particles per detector is expected. Simulation indicated, that in this case
underestimation is much less than $K_{2000} = 0.92$. It is connected with the fact that
conversion starts only after the first particle hit and at low density the effective
thickness of the shower front decreases. Probability of that the station doesn't operate
due to gap between operating of two separate detectors is more than 2~mcsec is 8.5\% and
it is lower by factor 3 than those due to Poisson fluctuations at this density. But
considering this circumstance together with steepness of the LDF at large distances we
restricted used effective area beyond the bound of array to obtain the intensity of the
largest showers.

Simulation showed no significant underestimation of particle density for distances up to
2000~m for density estimating system at the Yakutsk array. In the case of AGASA, when
input of RC--convector is constantly open, besides wide distribution, there is an
afterpulse contribution to density overestimation from delayed particles (probably
neutrons) together with casual additives from background muons. One can conclude from the
data in~\cite{bib7} that lagging neutrons can overstate the density by 1.37 already at
500~m and further. Background muons may cause distortions in wide range of axis distances.
If one such particle hits within last 10~mcsec of RC--circuit discharge, then resulting
density can be overestimated by factor 2 and more independently of the real density. The
effects mentioned above are excluded thanks to closing of converter's input in 2~mcsec.

\section{ENERGY SPECTRUM SUMMARY}

Events selection was provided as described in~\cite{bib3}. Showers with $\theta <
60^{\circ}$ were used. For determination of the intensity for showers with $\E > 4 \cdot
10^{19}$~eV, an extended area together efficient zone outside the array was used.

At standard procedure of axis determination, the Greisen--Linsley approximation of LDF
with parameters obtained before at the Yakutsk array is used~\cite{bib8}. It was shown
in~\cite{bib9}, that for showers with energy greater than $10^{19}$~eV this LDF badly
corresponded with experimental data at $R > 1000$~m from the axis. A modified
approximation was proposed:
\begin{eqnarray}
f(r) & \sim & \left(\frac{R}{\R}\right)^{-1.3} \cdot \left(1 + \frac{R}{\R}\right)^{-(b
              - 1.3)} \times \nonumber \\
     & &\times \left(1 + \frac{R}{2000}\right)^{-3.5}
	 \label{f1}
\end{eqnarray}
For showers with $\E > 2 \cdot 10^{19}$~eV parameter $b$ does not depend on energy but
depends on $\theta$. In this work we provided axes coordinates determination with this
adjusted LDF. As a result, in average estimated $S_{600}$ values increased: 10\% --- for
showers with axes lying within array area and 20\% on border.

For energy estimation we used adjusted formulas of $S_{300}$ and $S_{600}$~\cite{bib10}.
For upright showers ($\theta = 0^{\circ}$, depth $\X= 1020~\DepthUnit$), the following
relationships were obtained using calorimetric method:
\begin{eqnarray}
\E & = & (5.66 \pm 1.4) \cdot 10^{17} \times \nonumber \\
   & \times  & \left[\frac{S_{300}(0^{\circ})} {10}\right]^{0.94 \pm 0.02}~\textrm{eV,}
   \label{f2} \\
   &   & \nonumber \\
\E & = & (4.6 \pm 1.2) \cdot 10^{17} \times \nonumber \\
      & \times & [S_{600}(0^{\circ})]^{0.98 \pm 0.02}~\textrm{eV.} \label{f3}
\end{eqnarray}

Zenith-angular dependency was accepted in form of:
\newpage
\begin{eqnarray}
S(\theta) & = & S(0^{\circ}) \cdot
                \left[
			          (1 - \beta) \cdot \exp\left(\frac{\X - X}{\LambE}\right) +
					  \right. \nonumber \\
		&&		      \left. + \beta \cdot \exp\left(\frac{\X - X}{\LambM}\right)
			    \right]\textrm{,}\label{f4}
\end{eqnarray}
where $\LambE = 200~\AttenUnit$ --- the attenuation length for the soft
component (electrons), $\LambM = 1000~\AttenUnit$ --- the attenuation
length for the hard component (associated with muons), $\beta$ --- contribution from the
hard component to full response $S_{300}(0^{\circ})$ or $S_{600}(0^{\circ})$ at
$1020~\DepthUnit$ depth.

The following dependencies of parameters $\beta$ on $S_{300}$ and $S_{600}$ have been
carried out from experimental data:
\begin{eqnarray}
\beta_{300} & = & (0.368 \pm 0.021) \times \nonumber \\
            && \times \left[\frac{S_{300}(0^{\circ})} {10}\right]^{-0.185 \pm 0.02}
               \textrm{,} \label{f5} \\
& & \nonumber \\
\beta_{600} & = & (0.62 \pm 0.06) \times \nonumber \\
            && \times \left[S_{600}(0^{\circ})\right]^{-0.076 \pm 0.03}
               \textrm{.} \label{f6}
\end{eqnarray}

\begin{figure}[htb]
\centering
\includegraphics[width=5.7cm,angle=90]{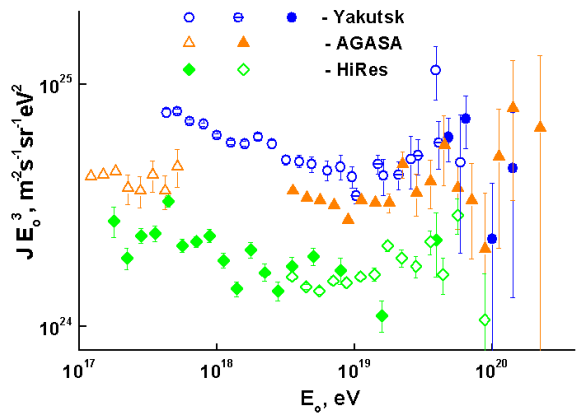}
\caption{Differential energy spectra according to the Yakutsk EAS array, AGASA~\cite{bib4}
and HiRes~\cite{bib5}.}
\label{fig1}
\end{figure}

On Fig.\ref{fig1} the differential energy spectra obtained at the Yakutsk array, AGASA and
HiRes are presented. Results obtained in different experiments correspond quite well in
shape but differ in intensity. The data from the Yakutsk array near $10^{19}$~eV are
higher by factor $\sim 2.5$ than HiRes data and $\sim 30$\% than AGASA's. This rather is
connected with the difference in estimation of the showers energy. For estimation of
energy in upright showers at AGASA the dependency between $S_{600}$ and $\E$ based upon
the model calculations~\cite{bib7} is used. Our formulas~(\ref{f2}) and (\ref{f3}) for
upright showers estimate the energy (30--40)\% higher at $\E \approxeq 5 \cdot 10^{17}$~eV
and (15--20)\% higher at $\E \approxeq 10^{19}$~eV than it follows from similar models.

\begin{table*}[htb]
\caption{The most energetic events detected with the Yakutsk array (sorted by energy)}
\label{tab}
\renewcommand{\tabcolsep}{1pc} 
\renewcommand{\arraystretch}{1.2} 
\begin{tabular}{@{}llllllll}
\hline
N&Date&Time, UT& $\theta^{\circ}$ & $\log \E$ & $\delta \E$ (\%) & $b^{\circ}$ & $l^{\circ}$ \\
\hline
1&02.18.04&22:20:38&47.7&20.16&42&16.3&140.2\\
2&05.07.89&22:03:00&58.7&20.14&46& 2.7&161.6\\
3&12.21.77&18:45:00&46.0&20.01&40&50.0&220.6\\
4&02.15.78&03:35:00& 9.6&19.99&32&15.5&102.0\\
\hline
\end{tabular}\\[2pt]
\end{table*}

In the region of ultra--high energies the results from Yakutsk and AGASA approach. There
are 4 events registered in Yakutsk with adjusted estimated energy exceeding $10^{19.9}$~
eV (see Table~\ref{tab}). Relative errors in energy estimation resulting from uncertainty
of the parameters in formulas and errors in determination of axis coordinates and zenith
angle in these individual showers amount from 32\% to 46\%. If the energy is reduced by
one standard error then it slightly exceeds the $10^{20}$~eV threshold only in one event.
Therefore the relic cutoff of the spectrum cannot be rejected based on Yakutsk EAS data.

Similar experimental errors are observed at AGASA. According to~\cite{bib10} their
averaged value is about 20\%. Taking into account this circumstance a conclusion was made
in~\cite{bib11} that yet there are too few events recorded to approve the spectrum cutoff
absence. Besides, estimations of the energy at AGASA depend on model conclusions. The
affects of observed densities mentioned above are also not considered yet.

The HiRes results are consistent with the GZK--cutoff of the spectrum, the AGASA and
Yakutsk data are inconsistent. But because of small statistics and errors in energy
estimation while it is impossible to final conclude about this problem. To solve this and
investigate of affinities of the particles with energies above GZK--cutoff, data with high
statistics and good accuracy in the energy estimation a necessary available.

\section*{Acknowledgements}
This work is supported by INTAS grant \textnumero03--51--5113 and Federal Agency of
Science and Innovations grant \textnumero748.2003.2.

\end{document}